\documentclass{osa-article}

\journal{osac}
\articletype{Research Article}
\begin{document}

\title{Switching between topological edge states in plasmonic systems using phase-change materials}

\author{Yin Huang,\authormark{1,5} Yuecheng Shen,\authormark{2,6} and Georgios Veronis\authormark{3,4}}

\address{\authormark{1}Department of Optoelectrics Information Science and Engineering, School of Physics and Electronics, Central South University, Changsha, Hunan 410012, China\\
\authormark{2}State Key Laboratory of Optoelectronic Materials and Technologies, School of Electronics and Information Technology, Sun Yat-Sen University, Guangzhou, Guangdong 510275, China}
\address{\authormark{3}School of Electrical Engineering and Computer Science, Louisiana State University, Baton Rouge, LA 70803, USA\\
\authormark{4}Center for Computation and Technology, Louisiana State University, Baton Rouge, LA 70803, USA\\
\authormark{5}yhuan15@csu.edu.cn \\
\authormark{6}shenyuecheng@mail.sysu.edu.cn}

% \email{\authormark{*}yhuan15@csu.edu.cn} %% email address is required



\begin{abstract}
We introduce non-Hermitian plasmonic waveguide-cavity structures based on the Aubry-Andre-Harper model to realize switching between right and left topological edge states (TESs) using the phase-change material Ge$_2$Sb$_2$Te$_5$ (GST). We show that switching between the crystalline and amorphous phases of GST leads to a shift of the dispersion relation of the optimized structure so that a right TES for the crystalline phase, and a left TES for the amorphous phase occur at the same frequency.
Thus, we realize switching between right and left TESs at that frequency by switching between the crystalline and amorphous phases of GST. Our results could be potentially important for developing compact reconfigurable topological photonic devices.
\end{abstract}

%%%%%%%%%%%%%%%%%%%%%%%%%%  body  %%%%%%%%%%%%%%%%%%%%%%%%%%
\section{Introduction}
Topological insulators possessing nontrivial topological states on their edge or surface have recently attracted considerable attention \cite{Bernevig:06,Hasan:10}. Topological edge states (TESs) associated with the existence of topological invariants in the bulk band structure have been reported in a variety of systems, such as graphene \cite{Hasan:10}, quantum Hall systems \cite{Ezawa} and the Su-Schrieffer-Heeger (SSH) model for polyacetylene \cite{Kane:13}. TESs are insensitive to disorder and can lead to field intensity enhancement \cite{Bernevig:06, Hasan:10}. However, conventional higher-dimensional topological systems require large footprints and are not easy to fabricate. Topological insulators which include synthetic dimensions have recently emerged as an attractive platform to realize higher-dimensional topological physics with lower dimensionality \cite{Yuan:18, Lustig:19}. Structures based on the Aubry-Andre-Harper (AAH) model have been shown to possess nontrivial topological states on the basis of tight-binding models \cite{Hofstadter:76, Lang:12}. The AAH model is the one-dimensional momentum-space projection of the two-dimensional integer quantum Hall system, and therefore exhibits similar topological properties \cite{Hofstadter:76}. It was recently demonstrated that TESs can emerge in non-Hermitian photonic and acoustic systems based on the AAH model \cite{Poshakinskiy:14, Zhu:18}. It was also demonstrated that TESs and exceptional points can be simultaneously realized in plasmonic waveguide-cavity systems based on the one-dimensional AAH model \cite{Huang:22}. 
TESs in one-dimensional photonic systems can enhance light-matter interactions and are therefore important for applications in sensing and nonlinear optics \cite{Budich:20, Liu:18, Guo:19}.

In a two-dimensional quantum Hall system, edge states at opposite edges propagate in opposite directions due to chirality \cite{Asboth}. Such states propagating in opposite directions can be excited at the same frequency. However, a consequence of chirality is that in structures based on the one-dimensional AAH model right and left TESs $\it cannot$ be excited at the same frequency \cite{Poshakinskiy:14, Zhu:18, Huang:22}. 
Switching between right and left TESs in such structures could be essential for building compact topological photonic devices, and for developing reconfigurable optical components.
This could be achieved by using materials with tunable optical properties such as phase-change materials.

Ge$_2$Sb$_2$Te$_5$ (GST) is a phase-change material with amorphous and crystalline phases \cite{Shportko:08}.
The amorphous phase is characterized by saturated covalent bonds, whereas in the crystalline phase resonant bonds are formed. Thus, GST can be switched between its amorphous and crystalline phases with a drastic change in its optical and electronic properties. The phase transition can be induced reversibly and rapidly by applying external electrical pulses, optical pulses or thermal annealing. The transition speed can be at subpicosecond timescales via femtosecond laser pulses \cite{Rude:16,Loke:12}. In addition, the phase change is nonvolatile and, therefore, power is consumed only during the phase transition process. No external power supply is required for GST to maintain its phase. GST is currently widely used in a variety of reconfigurable optical devices due to its switchable dielectric properties \cite{Schlich:15, Rude:15, Wang:16, Li:16, Mkhitaryan:17, Yoo:16, Huang:17}.

In this paper, we introduce non-Hermitian plasmonic waveguide-cavity structures based on the AAH model to realize switching between right and left TESs using the phase-change material GST. We show that switching between the crystalline and amorphous phases of GST leads to a shift of the dispersion relation of the optimized structure so that a right TES for the crystalline phase, and a left TES for the amorphous phase occur at the same frequency.
Thus, we realize switching between right and left TESs at that frequency by switching between the crystalline and amorphous phases of GST. We also find that the right and left TESs in the optimized structure are robust in the presence of disorder.

\section{Model}
\begin{figure}[htb]
\centering\includegraphics[width=9cm]{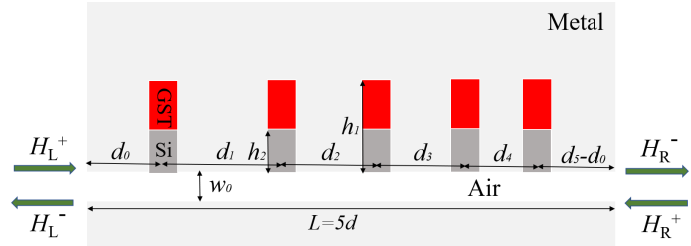}
\caption{Schematic of the compound unit cell of a periodic structure consisting of an MDM waveguide side coupled to five identical MDM stub resonators. The metal is silver and the stubs are filled with GST and silicon.}
\end{figure}
Analogous to the AAH model, we consider a periodic plasmonic structure with compound unit cells consisting of metal-dielectric-metal (MDM) stub resonators side-coupled to an MDM waveguide (Fig. 1). The compound unit cells consist of $N$ side-coupled MDM stub resonators. The distance between the $n$th and ($n+1$)th side-coupled stubs in the compound unit cell is modulated and given by \cite{Huang:22}
\begin{equation}
d_n=d\{1+\eta \cos[\frac{2\pi}{N} (n-1)+\phi]\}.
\end{equation}
Here, $d$ is the distance without modulation, and $\eta$ is the modulation strength. Figure 1 shows a compound unit cell with five side-coupled stubs ($N=5$). In addition, $\phi$ is an arbitrary phase. As $\phi$ is varied between $-\pi$ and $\pi$, the system characteristics continuously evolve, and for $\phi = \pi$ they are the same as for $\phi = -\pi$. Thus, the phase $\phi$ can be treated as a wavevector in an auxiliary direction, and one-dimensional problems can be mapped to two-dimensional integer quantum Hall effect problems with the Landau gauge characterized by nonzero Chern numbers \cite{Lang:12}.

The properties of the compound unit cell of Fig. 1 can be described by the transfer matrix $\mathbf{M}$ defined by the following equation
\begin{equation}
\left[
\begin{array}{c}
H_R^-\\
H_R^+\\
\end{array}
\right]
=
\mathbf{M}\left[
\begin{array}{c}
H_L^+\\
H_L^-\\
\end{array}
\right]=\left[
\begin{array}{c c}
M_{11} & M_{12} \\
M_{21} & M_{22} \\
\end{array}
\right]\left[
\begin{array}{c}
H_L^+\\
H_L^-\\
\end{array}
\right].
\end{equation}
Here $H^+_L$ , and $H^-_L$ are the complex magnetic field amplitudes of the incoming and outgoing modes at the left port, respectively. Similarly, $H^+_R$ , and $H^-_R$ are the complex magnetic field amplitudes
of the incoming and outgoing modes at the right port, respectively (Fig. 1). The transfer matrix $\mathbf{M}$ can be calculated by
\begin{equation}
\mathbf{M}=\mathbf{M}_6\mathbf{M}_s\mathbf{M}_5\mathbf{M}_s\mathbf{M}_4\mathbf{M}_s\mathbf{M}_3\mathbf{M}_s\mathbf{M}_2\mathbf{M}_s\mathbf{M}_1,
\end{equation}
where
$\mathbf{M}_s=
\left[
\begin{array}{c c}
t_s-\frac{r_s^2}{t_s} & \frac{r_s}{t_s} \\
-\frac{r_s}{t_s} & \frac{1}{t_s} \\
\end{array}
\right]$ is the transfer matrix of a system consisting of an MDM waveguide side-coupled to a stub, while $r_s$ and $t_s$ are the complex reflection and transmission coefficients of the system, respectively, which can be obtained using temporal coupled mode theory (CMT) \cite{Joannopoulos}.
In addition, $\mathbf{M}_i=\left[
\begin{array}{c c}
e^{-\gamma L_i} & 0 \\
0 & e^{\gamma L_i} \\
\end{array}
\right], i=1, 2, 3, 4, 5, 6$, where $L_1=d_0$, $L_2=d_1$, $L_3=d_2$, $L_4=d_3$, $L_5=d_4$, $L_6=d_5-d_0$, $d_0$ is an arbitrary distance with $d_0<d_5$, and $\gamma$ is the complex propagation constant of the propagating fundamental mode of the MDM waveguide.
Imposing the Bloch boundary condition, the right and left reflection coefficients for the semi-infinite structure can be obtained using the transfer matrix
\begin{equation}
r_{\infty,r}=\frac{e^{-jkL}-M_{22}}{M_{21}}, r_{\infty,l}=\frac{e^{jkL}-M_{11}}{M_{12}}.
\end{equation}
Note that, in the lossless case, the left and right reflection coefficients are equal.

In addition, the band structure of the periodic plasmonic structure with the compound unit cell of Fig. 1 can be calculated using the transfer matrix $\mathbf{M}$ of the unit cell and the Bloch boundary conditions $H^-_R=e^{jkL}H^+_L$, $H^+_R=e^{jkL}H^-_L$ by
\begin{equation}
\bigg|\mathbf{M}- \left[
\begin{array}{c c}
e^{-jkL} & 0 \\
0 & e^{jkL} \\
\end{array}
\right]\bigg|=0,
\end{equation}
where $L=5d$ is the overall length of the compound unit cell (Fig. 1), and $k$ is the Bloch wavevector.

\section{Results}
In this section, we design plasmonic waveguide-cavity structures with a phase-change material, to realize switching between right and left TESs. We use experimental data for the frequency-dependent dielectric constants of all materials, including GST in its amorphous (aGST) and crystalline (cGST) phases \cite{Mkhitaryan:17}. In Figs. 2(a) and 2(b) we show the real and imaginary parts, respectively, of the refractive index of aGST and cGST as a function of wavelength.
\subsection{Right topological edge states for GST in its crystalline phase}
\begin{figure}[htb]
\centering\includegraphics[width=11cm]{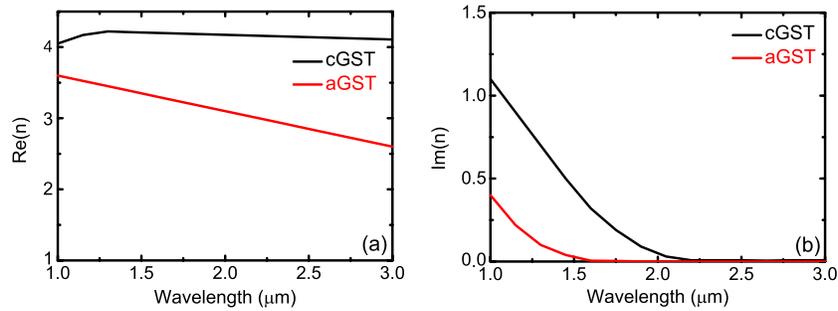}
\caption{(a), (b) Real and imaginary parts, respectively, of the refractive index of aGST and cGST as a function of wavelength.}
\end{figure}

We first consider a periodic plasmonic structure without modulation [$\eta=0$ in Eq. (1) so that $d_1=d_2=d_3=d_4=d_5=d$]. We choose $d=550$ nm, $h_1=100$ nm, $h_2=40$ nm, and $w_0=50$ nm (Fig. 1). The metal is silver and the stubs are filled with GST in its crystalline phase (cGST) and silicon (Fig. 1). We first assume that all materials are lossless. However, material loss will be included later. In Fig. 3(a), we show the dispersion relation of the structure without modulation calculated using the finite-difference frequency-domain (FDFD) method (black squares). The middle of the three bands shown corresponds to a mode with slow group velocity. We also consider a structure in which the distances between adjacent side-coupled stubs are modulated as in Eq. (1) with $N=5$, $\eta=0.4$, and $\phi=0$. The compound unit cell has overall length $L=5d$ (Fig. 1). All three bands of the structure without modulation shown in Fig. 3(a) (black squares) split due to band mixing for the structure with modulation [circles in Fig. 3(a)]. In particular, the middle band with slow group velocity of the structure without modulation [black squares in Fig. 3(a)] splits into five bands for the structure with modulation [red circles in Fig. 3(a)]. Figure 3(b) also shows the dispersion relation of the structure with modulation calculated using CMT [Eq. (5)] (red circles). We observe that there is very good agreement between the CMT results and the exact results obtained using FDFD.

\begin{figure}[htb]
\centering\includegraphics[width=11cm]{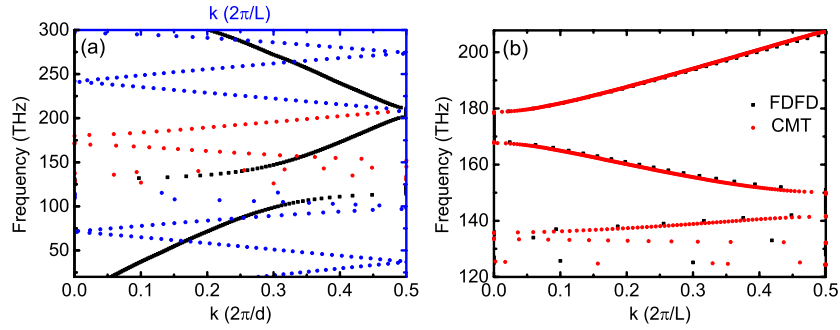}
\caption{(a) Dispersion relation of the periodic structure without modulation calculated using FDFD (black squares). Results are shown for $d_1=d_2=d_3=d_4=d_5=d=550$ nm, $h_1=100$ nm, $h_2=40$ nm, $w_0=50$ nm (Fig. 1), and $\eta=0$ [Eq. (1)]. The stubs are filled with GST in its crystalline phase (cGST) and silicon, and the metal is silver. Here, we assume that cGST and silver are lossless. Also shown is the dispersion relation of a periodic structure with the compound
unit cell of Fig. 1, in which the distances between adjacent side-coupled stubs are modulated as in Eq. (1), calculated using FDFD (circles). 
The middle band of the structure without modulation (black squares) splits into five bands for the structure with modulation (red circles).
The upper and lower bands of the structure without modulation (black squares) split into several bands for the structure with modulation (blue circles).
Results are shown for $d_1=d(1+\eta \cos\phi)$, $d_2=d[1+\eta \cos(\frac{2\pi}{5}+\phi)]$, $d_3=d[1+\eta \cos(\frac{4\pi}{5}+\phi)]$, $d_4=d[1+\eta \cos(\frac{6\pi}{5}+\phi)]$, $d_5=d[1+\eta \cos(\frac{8\pi}{5}+\phi)]$, $\eta=0.4$, and $\phi=0$. All other parameters are the same as in the structure without modulation. (b) Zoomed-in view of the dispersion relation in a narrower frequency range for the periodic structure in which the distances between adjacent side-coupled stubs are modulated. The dispersion relation is calculated using FDFD (black squares) and CMT (red circles). All other parameters are as in (a).}
\end{figure}

Figure 4(a) shows the projected band structure of the periodic structure with the compound unit cell of Fig. 1 for $\eta=0.4$ and $N=5$ as a function of the phase $\phi$ [Eq. (1)], calculated with CMT. For a fixed $\phi$, we have five bands (yellow regions) in the frequency range of interest. As an example, for $\phi=0$ the system supports five bands with frequencies ranging from 125 to 126 THz, 133 to 135 THz, 136 to 142 THz, 150 to 168 THz, and 179 to 208 THz [Figs. 3(b) and 4]. As the phase $\phi$ varies from $-\pi$ to $\pi$, the Bloch eigenstates of the system are a function of $k$ and $\phi$. Each band is characterized by a Chern number. This Chern number can be deduced from the winding numbers of the above-lying and below-lying bandgaps using the phase of the reflection coefficient for the semi-infinite structure $r_\infty$ \cite{Zhu:18,Poshakinskiy:15}. The absolute value of the reflection coefficient $|r_{\infty}|$ for frequencies lying inside the bandgaps is 1. Thus, the reflection coefficient for the semi-infinite structure for frequencies lying inside the bandgaps is $r_\infty=1e^{j\theta}$, where $\theta$ is the phase of the reflection coefficient. The winding number, which is the topological invariant of the bandgap, can be calculated using \cite{Asboth}
\begin{equation}
w=\frac{1}{2\pi j}\int_0^{2\pi}\frac{\partial \ln[r_\infty(\phi)]}{\partial \phi}d\phi=\frac{1}{2\pi }\int_0^{2\pi}\partial\theta(\phi).
\end{equation}

Figure 4(b) shows the phase $\theta$ of the reflection coefficient $r_{\infty,r}$ when the waveguide mode is incident from the right onto the semi-infinite structure. The extra phase that the reflection coefficient accumulates when $\phi$ varies from $-\pi$ to $\pi$ is 0, $2\pi$, $4\pi$, $-4\pi$, $-2\pi$, and 0 for the first, second, third, fourth, fifth and sixth bandgap, respectively, in the frequency range of interest. Thus, the winding numbers of these six bandgaps are 0, 1, 2, -2, -1 and 0 [Eq. (6)] [Fig. 4(b)]. Thus, since the Chern number of a band is equal to the winding number of the above-lying bandgap minus the winding number of below-lying bandgap \cite{Poshakinskiy:15}, the Chern numbers of the five bands in the frequency range of interest are 1, 1, -4, 1, and 1 [Fig. 4(a)]. The sum of Chern numbers over all bands is zero \cite{Wang:09}. Different Chern numbers in adjacent bands indicate the existence of TESs in the third and fourth bandgaps [Fig. 4(a)] based on the bulk-boundary correspondence \cite{Asboth}.

\begin{figure}[htb]
\centering\includegraphics[width=11.5cm]{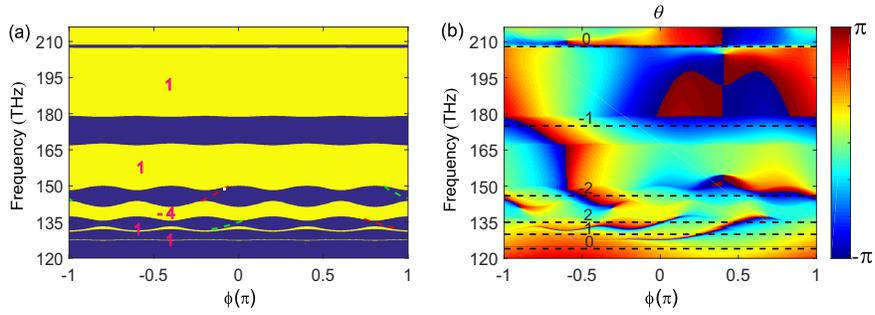}
\caption{(a) Projected dispersion relation of the periodic structure with the compound unit cell of Fig. 1 as a function of $\phi$ calculated with CMT with GST in its crystalline phase (cGST). The yellow regions correspond to the bands separated by the bandgaps (blue regions). Also shown is the Chern number of each band.
Green (red) dashed lines correspond to left (right) edge states.
All other parameters are as in Fig. 3(b). (b) The phase $\theta$ of the reflection coefficient $r_{\infty,r}$ as a function of $\phi$, when the waveguide mode is incident from the right onto the semi-infinite structure. Also shown is the winding number of each bandgap. All parameters are as in (a).}
\end{figure}

The left (right) TESs can be  extracted from the minima in the reflection spectra of the semi-infinite structure $|r_{\infty,l(r)}|^2$ when material loss is included  \cite{Poshakinskiy:14,Zhu:18,Huang:22}. The underlying physical mechanism behind this can be explained as follows: when light with the eigenfrequency of the edge state is incident on the structure, the edge state is excited and light is deeply trapped in the edge region. If loss is introduced into the system, the trapped light is absorbed, and thus $r_{\infty,l(r)}(\omega)$ is minimized. Figure 4(a) shows that the structure possesses two chiral edge states in each of the third and fourth bandgaps when we take into account the material loss of silver and cGST. Green dashed lines correspond to left edge states, while red dashed lines correspond to right edge states. Chirality of edge states is a property of two-dimensional integer quantum Hall systems. AAH systems are the one-dimensional momentum-space projection of two-dimensional integer quantum Hall systems and exhibit non-trivial topological properties \cite{Lang:12}. 
The modulation distance in Eq. (1) provides an effective gauge magnetic field, and the phase $\phi$ accounts for the momentum along the geometrical dimension that was lost when moving from a two-dimensional to a one-dimensional system. A consequence of chirality in the one-dimensional AAH model is that in a given structure with a specific phase $\phi$, right and left edge states cannot be excited at the same frequency. Thus, right and left TESs at the same frequency correspond to different structures with different $\phi$. For example, left and right TESs are obtained at $f=$ 134.9 THz for $\phi=0$ ($d_1=1.40d, d_2=1.12d, d_3=0.68d, d_4=0.68d, d_5=1.12d$) and $\phi=0.8\pi$ ($d_1=0.68d, d_2=0.68d, d_3=1.12d, d_4=1.40d, d_5=1.12d$), respectively. 

\subsection{Left topological edge states for GST in its amorphous phase}
\begin{figure}[htb]
\centering\includegraphics[width=11.5cm]{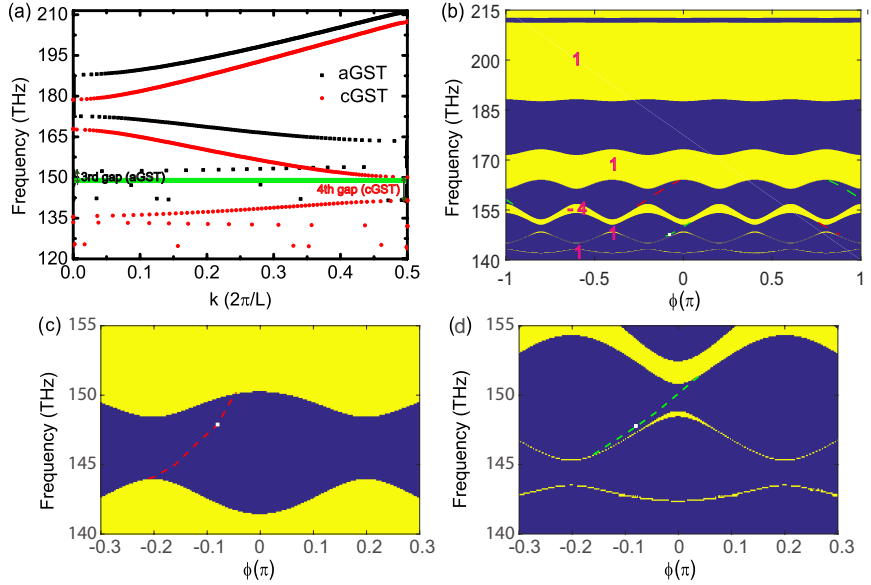}
\caption{(a) Dispersion relation of the periodic structure with the compound unit cell of Fig. 1, in which the distances between adjacent side-coupled stubs are modulated as in Eq. (1), calculated using CMT for GST in its crystalline (red circles) and its amorphous (black squares) phases for $\phi=-0.08 \pi$. All other parameters are as in Fig. 3(b). We also indicate with arrows the frequency range of the fourth and third bandgap for the cGST and aGST cases, respectively. 
The overlapping frequency range of these bandgaps is highlighted in green.
(b) Projected dispersion relation of the periodic structure with the compound unit cell of Fig. 1 as a function of $\phi$ calculated using CMT with GST in its amorphous phase (aGST). 
The yellow regions correspond to the bands separated by the bandgaps (blue regions). Also shown is the Chern number of each band. Green (red) dashed lines correspond to left (right) edge states.
All other parameters are as in (a). (c) and (d) 
Zoomed-in views of the projected dispersion relation in a narrower frequency and $\phi$ range for GST in its crystalline (cGST) and amorphous (aGST) phases, respectively. All other parameters are as in (b).}
\end{figure}

In Fig. 4(a), we observe that for specific values of $\phi$ (e.g., $\phi=-0.08 \pi$) the corresponding structures support both right and left TESs at different frequencies. When this happens, the right and left TESs are located at different bandgaps. 
This suggests that, in order to achieve switching between right and left TESs in the structure of Fig. 1, switching between the crystalline and amorphous phases of GST should lead to a shift of the dispersion relation of the structure so that a right TES for one phase, and a left TES for the other phase occur at the same frequency.
Thus, if the following two requirements are satisfied, we can realize switching between right and left TESs. First, the fourth bandgap for GST in its crystalline phase (cGST) must overlap with the third bandgap for GST in its amorphous phase (aGST). Second, the spectra $|r_{\infty,r(l)}|^2$ for the cGST case and $|r_{\infty,l(r)}|^2$ for the aGST case must both be minimized at the same frequency, when loss is introduced into the semi-infinite structure. To fulfill these two requirements, we use CMT to optimize the distance $d$, length $h_2$ (Fig. 1), and phase $\phi$ [Eq. (1)]. Using this approach, we find that for $d = 550$ nm, $h_2 = 40$ nm, and $\phi=-0.08 \pi$ the system supports right and left TESs for GST in its crystalline (cGST) and amorphous (aGST) phases, respectively, at $f=148$ THz. 
 Fig. 5(a) shows the dispersion relation of the optimized structure calculated using CMT for GST in its crystalline and amorphous phases (shown with red circles and black squares, respectively). We observe that indeed the fourth bandgap for GST in its crystalline phase (cGST) partially overlaps with the third bandgap for GST in its amorphous phase (aGST). The overlapping frequency range is highlighted in green.
As shown in Fig. 4(a), a right TES exists for $\phi=-0.08 \pi$ at $f=148$ THz for GST in its crystalline phase (white square on the red dashed line). In addition, Fig. 5(b) shows that a left TES exists for the same structure ($\phi=-0.08 \pi$) at the same frequency ($f=148$ THz) when GST is in its amorphous phase (white square on the green dashed line). Interestingly, the topological properties remain the same when GST is switched between its crystalline and amorphous phases. Even though the frequency range of the five bands of interest is shifted when switching between cGST and aGST, their Chern numbers (1, 1, -4, 1, and 1) remain the same [Figs. 4(a) and 5(b)]. Figures 5(c) and 5(d) show zoomed-in views of the projected dispersion relation in a narrower frequency and $\phi$ range for GST in its crystalline (cGST) and amorphous (aGST) phases, respectively. These clearly demonstrate that the right TES for cGST [Fig. 5(c)] exists for the same structure ($\phi=-0.08 \pi$) and at the same frequency ($f=148$ THz) as the left TES for aGST [Fig. 5(d)].

\begin{figure}[htb]
\centering\includegraphics[width=10.5cm]{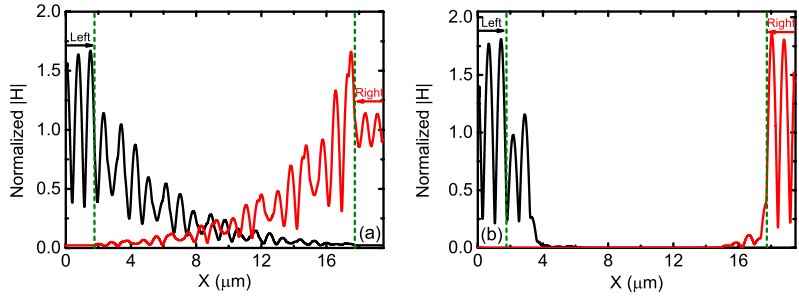}
\caption{(a) and (b) Profile of the magnetic field amplitude in the middle of the MDM waveguide, normalized with respect to the field amplitude of the incident waveguide mode in the middle of the waveguide, for GST in its crystalline (cGST) and amorphous (aGST) phases, respectively, calculated using FDFD.
Results are shown for the mode incident from the left (black solid line) and right (red solid line) onto the six-unit-cell structure at f = 148 THz.
The vertical dashed lines indicate the boundaries between the six-unit-cell structure and the MDM waveguide. All other parameters are as in Fig. 5(a).}
\end{figure}

To further confirm the existence of the right TES for cGST and the left TES for aGST, in Fig. 6 we show the normalized magnetic field distributions at $f=148$ THz calculated using FDFD for a finite structure consisting of six optimized compound unit cells as in Fig. 1. 
We found that six unit cells are sufficient for the finite structure to form bandgaps and for TESs to appear within these bandgaps \cite{Zhu:18, Huang:22}. The material loss in silver and GST is included here. 
When GST is in its crystalline phase [Fig. 6(a)] and the mode is incident from the right (red solid line), the field inside the structure is enhanced at the right edge demonstrating the existence of the right TES. 
When GST is in its amorphous phase [Fig. 6(b)] and the mode is incident from the left (black solid line), the field inside the structure first increases close to the left edge and then rapidly decays. This field distribution demonstrates the existence of the left TES.

In addition, we expect the TESs to be robust in the presence of disorder, as long as the disorder is not strong enough to close the bandgaps \cite{Huang:22}. To test this hypothesis, we randomly varied the distances between adjacent stubs in the six-unit-cell structure with variations which are uniformly distributed over the interval ($-40$ nm, 40 nm). We found that the magnetic field distributions of the disordered structures confirm the robustness of the TESs in the presence of disorder.

\section{Conclusions}
In this paper, we designed non-Hermitian plasmonic waveguide-cavity structures based on the AAH model with the phase-change material GST to realize switching between right and left TESs. We used the transfer matrix method and CMT to account for the behavior of the proposed structures.
We found that CMT results are in very good agreement with the exact results obtained using FDFD.
We showed that switching between the crystalline and amorphous phases of GST leads to a shift of the dispersion relation of the optimized structure so that a right TES for the crystalline phase (cGST), and a left TES for the amorphous phase (aGST) occur at the same frequency.
Thus, we realize switching between right and left TESs at that frequency by switching between cGST and aGST. Interestingly, even though the dispersion relation of the optimized structure is shifted when switching between cGST and aGST, the Chern numbers of the bands remain the same. We confirmed the existence of the right TES for cGST and the left TES for aGST in the magnetic field distributions of a finite structure consisting of six optimized compound unit cells. We also found that the TESs in the optimized structure are robust in the presence of disorder.

As final remarks, the optimized compound unit cell (Fig. 1) includes five side-coupled MDM stub resonators [$N=5$ in Eq. (1)]. We found that smaller number of side-coupled stubs ($N<5$) leads to sharper slope of the edge states in the dispersion diagram of the structure for GST in its crystalline (cGST) and amorphous (aGST) phases. Due to this sharper slope for $N<5$, it is impossible to realize a right TES for cGST and a left TES for aGST at the same frequency.
Although here we focused on plasmonic structures, the concept of using phase change materials to switch between different types of TESs could also be applied in other nanophotonic structures.
We also note that switching between right and left TESs could be realized in three-dimensional plasmonic waveguide-cavity systems based on plasmonic coaxial waveguides \cite{Shin:13, Mahigir:15}.

\section*{Funding}
National Natural Science Foundation of China (61605252); National Key Research and Development Program of China (2019YFA0706301); National Natural Science Foundation of China (12004446).

\section*{Disclosures}
The authors declare no conflicts of interest.

\section*{Data Availability}
Data underlying the results presented in this paper are not publicly available at this time but may be obtained from the authors upon reasonable request.

%%%%%%%%%%%%%%%%%%%%%%% References %%%%%%%%%%%%%%%%%%%%%%%%%

%Add references with BibTeX or manually.
%\cite{Zhang:14,OSA,FORSTER2007,Dean2006}
%
%%%%%%%%%%% If using BibTeX:
%\bibliography{sample}
%
%%%%%%%%% If preparing manually:

\end{document}